\documentclass[conference]{IEEEtran}
\IEEEoverridecommandlockouts
\usepackage[nolist,nohyperlinks]{acronym}
\def\BibTeX{{\rm B\kern-.05em{\sc i\kern-.025em b}\kern-.08em
    T\kern-.1667em\lower.7ex\hbox{E}\kern-.125emX}}

\newcommand*{\mb}[1]{\mathbf{#1}}

\usepackage{enumerate}
\usepackage{algorithmicx, algpseudocode, algorithm}
\algrenewcommand\algorithmicrequire{\textbf{Iteration 0: }}
\algrenewcommand\algorithmicensure{\textbf{Iteration $k$: }}

\usepackage{caption}
\usepackage{graphics}
\usepackage{xparse}
\usepackage{csvsimple}
\usepackage{balance}
\usepackage{ifthen}

\usepackage{hyperref}
\usepackage[nospace]{cite}
\usepackage{amsmath,amssymb,amsfonts}
\usepackage{graphicx}
\usepackage{textcomp}
\usepackage{xcolor}
\usepackage{soul}
\usepackage{listings}
\usepackage{multirow}
\usepackage[export]{adjustbox}
\usepackage{mathabx}
\usepackage{array}
\usepackage{mdwmath}
\usepackage{mdwtab}
\usepackage{eqparbox}
\usepackage{url}
\usepackage{color}
\usepackage{bm}
\usepackage{booktabs}

\usepackage[tableposition=top]{caption} 


\floatname{algorithm}{Proposed Algorithm}

\ifCLASSOPTIONcompsoc
 \usepackage[caption=false,font=normalsize,labelfont=sf,textfont=sf]{subfig}
\else
 \usepackage[caption=false,font=footnotesize]{subfig}
\fi
\usepackage{stfloats}

\newcommand{\version}{draft}

\usepackage{xparse}

\ExplSyntaxOn
\NewExpandableDocumentCommand{\ifstringsequalTF}{mmmm}
 {
  \str_if_eq:eeTF { #1 } { #2 } { #3 } { #4 }
 }
\NewExpandableDocumentCommand{\stringcase}{mO{}m}
 {
  \str_case_e:nnF { #1 } { #3 } { #2 }
 }
\ExplSyntaxOff

\ifstringsequalTF{\version}{draft}
  {
    \newcommand{\added}[1]{{\color{blue}#1}}
    \newcommand{\deleted}[1]{{\color{red}\st{#1}}}
    \newcommand{\noted}[1]{{\color{orange}#1}}
  }
  {
    \ifstringsequalTF{\version}{final} {
        \newcommand{\added}[1]{#1}
        \newcommand{\deleted}[1]{}
        \newcommand{\noted}[1]{}
    } 
    {
      \ifstringsequalTF{\version}{old} {
        \newcommand{\added}[1]{}
        \newcommand{\deleted}[1]{{\color{red}#1}}
        \newcommand{\noted}[1]{}
      }
      {
        \newcommand{\added}[1]{{#1}}
        \newcommand{\deleted}[1]{}
        \newcommand{\noted}[1]{}
      }
    }    
  }

\begin{document}
\def\thetitle{Deep Learning Based Multi-Step Channel Prediction for Adaptive Underwater Acoustic OFDM Systems}
\title{\thetitle}
\vspace{-5mm}
\author{
\IEEEauthorblockN{
    Tian Tian \IEEEauthorrefmark{1}, 
    Ying Zhang \IEEEauthorrefmark{2},  
    Agastya Raj \IEEEauthorrefmark{1},
    Fei-Yun Wu \IEEEauthorrefmark{4},
    Marco Ruffini \IEEEauthorrefmark{1}
}

\IEEEauthorblockA{\IEEEauthorrefmark{1}
    IRIS Research Group, ADAPT Center, School of Computer Science and Statistics, Trinity College Dublin, Ireland}
\IEEEauthorblockA{\IEEEauthorrefmark{2}
 College of Oceanography, Hohai University, Nanjing, Jiangsu, China}
\IEEEauthorblockA{\IEEEauthorrefmark{4}
    Navigation College, Jimei University, Xiamen, Fujian, China}
    \vspace{-11mm}

}
\maketitle

\begin{abstract}
We develop an adaptive OFDM framework for underwater acoustic communications based on PatchCSI-T, a Transformer-based multistep channel prediction model with feature-independent modeling and parameter sharing. Combined with a greedy adaptive modulation and power allocation scheme, the proposed approach enables accurate, low-latency CSI forecasting and improves end-to-end BER and spectral efficiency on real-world UWA channel datasets.
\end{abstract}
\section{Introduction}\label{sec:intro}
Acoustic propagation in the ocean is governed by frequency-dependent attenuation, severe time-varying multipath fading, motion-induced Doppler shifting and long propagation delays that are more pronounced due to the low sound speed \cite{stojanovicUnderwaterAcousticCommunication2009a}. These characteristics make timely \ac{csi} acquisition difficult as round-trip propagation delays over kilometer-scale links can reach seconds, rendering feedback-based link adaptation impractical when the channel coherence time is of similar order.

Channel prediction mitigates feedback latency by enabling proactive link adaptation without waiting for round-trip \ac{csi} feedback. Accurate multi-step-ahead \ac{csi} prediction allows adaptive communication systems to operate on predicted rather than outdated channel conditions, thereby improving spectral efficiency and reducing \ac{ber}\cite{huangEfficiencyEnhancementUnderwater2020}. Conventional adaptive filtering methods, such as \ac{rls}, have been widely used for \ac{uwa} channel estimation and prediction\cite{radosevicAdaptiveOFDMModulation2014}. However, these methods often struggle to capture the nonlinear and rapidly time-varying characteristics of \ac{uwa} channels.

Recently, \ac{dl}-based methods have shown potential for \ac{uwa} channel prediction. Existing studies include hybrid CNN–RNN architectures for \ac{csi} prediction in adaptive downlink OFDMA systems \cite{liuChannelStateInformation2021a},  attention-based recurrent models for capturing long-range temporal dependencies in time-varying channels \cite{zhuDeepLearningPrediction2023}, bidirectional GRU-based models for \ac{uwa} MIMO systems \cite{huChannelPredictionUsing2023}, and \ac{mtl} frameworks for high-dimensional \ac{cir} prediction \cite{tian}. However, current \ac{dl}-based channel prediction methods still face important limitations. RNN-based architectures that rely on autoregressive multi-step prediction suffer from error accumulation, as prediction errors propagate and amplify over long horizons. In addition, architectures that employ cross-dimension feature mixing to capture inter-subcarrier or inter-tap correlations are prone to overfitting when training data are limited, which is a common constraint in \ac{uwa} applications due to the high cost of at-sea measurements.

Motivated by the limitations of existing methods, this paper proposes an adaptive OFDM framework with two key components: (1) PatchCSI-T, a Transformer-based multi-step channel prediction model inspired by time-series forecasting model PatchTST \cite{nieTimeSeriesWorth2023}, which leverages patching, feature-independent modeling, and parameter sharing to reduce attention complexity and mitigate overfitting; and (2) a greedy adaptive modulation and power allocation scheme that jointly selects the modulation order and transmit power to satisfy a target \ac{ber} while maximizing spectral efficiency. Experiments on field-measured \ac{uwa} channel datasets show that PatchCSI-T outperforms state-of-the-art baselines in both prediction accuracy and inference efficiency. When integrated into the adaptive OFDM system, the proposed framework further improves end-to-end \ac{ber} and spectral efficiency.
\begin{figure*}[h]
    \centering
    \includegraphics[width=0.98\textwidth]{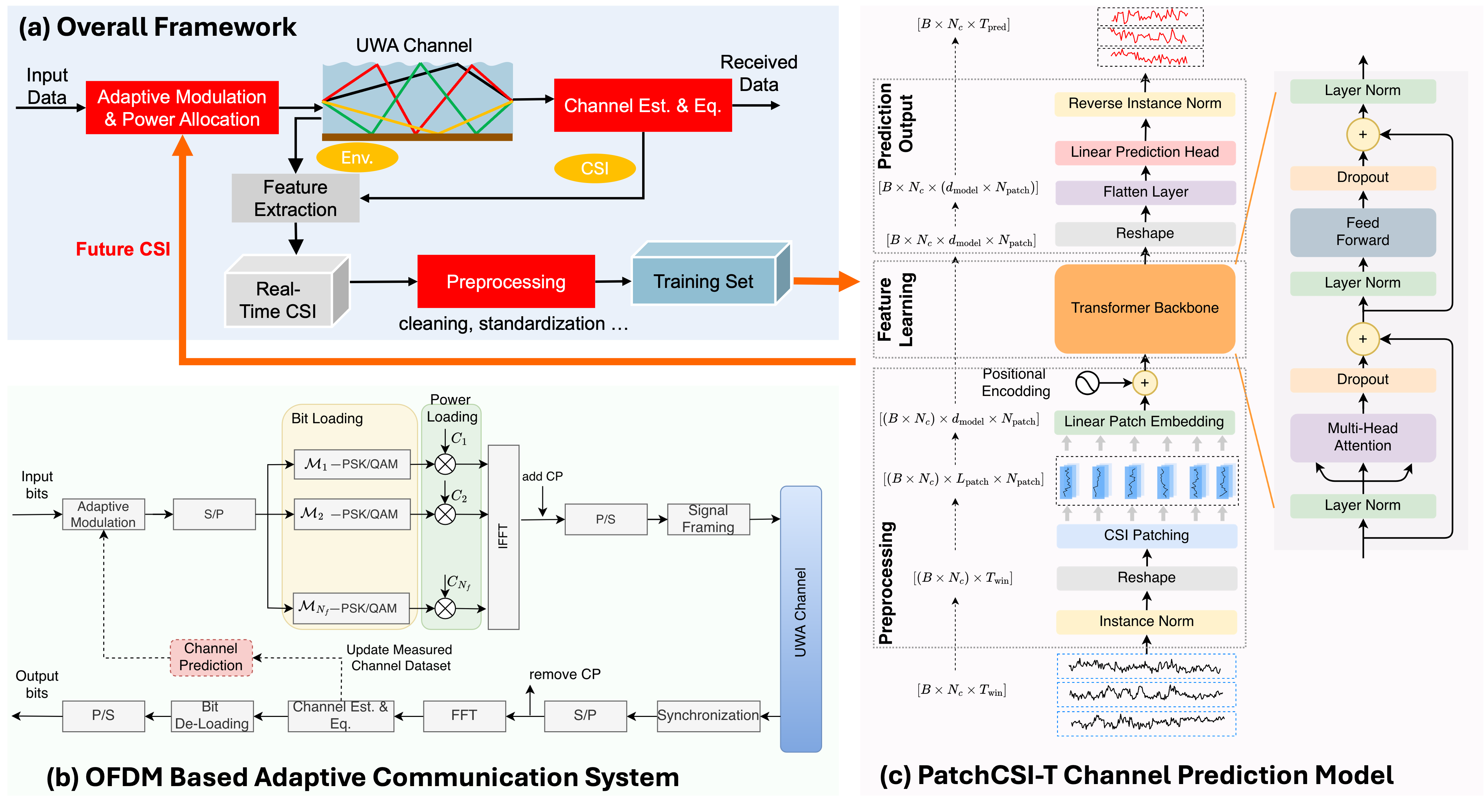}
    \caption{System framework and channel prediction model.}
    \label{fig:framework}
\end{figure*}
\section{System Model and Proposed Framework}
The overall system framework and the structure of the PatchCSI-T channel prediction model are shown in Fig. \ref{fig:framework}. The transmitter adaptively maps input bits onto $\mathcal{M}$-ary symbols according to the predicted CSI, applies per-subcarrier power loading, and converts the signal to the time domain via IFFT. At the receiver, after frame synchronization and FFT processing, channel estimation and equalization are performed, followed by symbol demapping to recover the transmitted bits.

\subsection{Problem Formulation}
Consider an OFDM system with $N_f$ subcarriers. Let $G_i$ denote the complex channel gain on subcarrier $i$ and $C_i$ the power allocated to that subcarrier. Each subcarrier employs an $M$-ary modulation, with modulation order $\mathcal{M}_i$ corresponding to $b_i = \log_2 \mathcal{M}_i$ bits per symbol. The adaptive modulation and power allocation problem seeks to maximize throughput subject to a total power budget $C^{\mathrm{budget}}$, a target BER $P_b^{\mathrm{target}}$, and a per-subcarrier modulation constraint:
{\small{
\begin{equation}
\label{eq.amc_obj}
\max \sum_{i=1}^{N_f} b_i \ \text{s.t.} \
\sum_{i=1}^{N_f} C_i \le C^{\text{budget}},\ 
\frac{1}{N_f}\sum_{i=1}^{N_f} P_{b,i} \leq P_b^{\text{target}},\ 
b_i \le b^{\max}.
\end{equation}}}
For a given modulation order $\mathcal{M}_i$ and per-subcarrier \ac{snr} $\gamma_i = C_i |G_i|^2 / N_0$, the \ac{ber} of $\mathcal{M}_i$-ary modulation can be approximated as \cite{radosevicAdaptiveOFDMModulation2014}:
\begin{equation}
    P_{b,i}(\gamma_i, \mathcal{M}_{i})\approx0.2\exp\left(-q(\mathcal{M}_i)\gamma_i\right)
\end{equation}
where $q(\mathcal{M}_i)$ is a modulation-dependent constant with values $q(\mathcal{M}_i)=1.1,0.55,0.25,0.12$ for $\mathcal{M}_i=2,4,8,16$, respectively. Since the system operates based on predicted rather than perfect \ac{csi}, the prediction error is modeled as $e_i=\hat{G}_i-G_i$, where $\hat{G}_i$ denotes the predicted channel gain and $e_i\sim\mathcal{CN}(0,\sigma_e^2)$. Under the Gaussian error model, the conditional expected \ac{ber} on subcarrier $i$ is
\begin{equation}
\begin{aligned}
    \label{eq.chap5_prob_ber}
    \mathbb{E}[P_{b,i}\vert\hat{G}_i]=
    0.2\frac{\exp\left(-\frac{|\hat{G}_i|^2}{\sigma_e^2}\left(1-\frac{1}{1+q(\mathcal{M}_i)\frac{C_i}{{N}_0}\sigma_e^2}\right)\right)}{1+q(\mathcal{M}_i)\frac{C_i}{{N}_0}\sigma_e^2}
\end{aligned}
\end{equation}
Imposing the constraint $\mathbb{E}[P_{b,i}\vert\hat{G}_i]\leq P_b^{\text{target}}$ and solving for $C_i$, a closed-form minimum power threshold can be obtained in terms of the Lambert-W function $W_0(\cdot)$ \cite{radosevicAdaptiveOFDMModulation2014}:
{\small{
\begin{equation}
\label{eq.chap5_power_thre}
\begin{split}
    C_{\text{thre},i}(\mathcal{M}_i,\hat{G}_i,P_b^{\text{target}})=& \frac{{N}_0}{q(\mathcal{M}_i)\sigma_e^2}
    \left[ \frac{|\hat{G}_i|^2}{\sigma_e^2} \right. \\
    &\left. W_0\left( \frac{P_b^{\text{target}}}{0.2} \exp\left( \frac{|\hat{G}_i|^2}{\sigma_e^2} \right) \frac{|\hat{G}_i|^2}{\sigma_e^2} \right)^{-1} - 1 \right].
\end{split}
\end{equation}}}
This threshold represents the minimum transmit power required on subcarrier $i$ to satisfy target \ac{ber} for a chosen modulation order $\mathcal{M}_i$, accounting for \ac{csi} prediction uncertainty.

\subsection{Greedy Bit / Power Allocation}
Directly solving \eqref{eq.amc_obj} is computationally expensive. We therefore employ a low-complexity greedy algorithm that iteratively adjusts per-subcarrier modulation orders using the power threshold \eqref{eq.chap5_power_thre}. At each iteration $n$, the most efficient bit upgrade is identified by:
\begin{equation}\label{eq.amc_sub}
\begin{aligned}
     &i^{\star}=\underset{1\leq i \leq N_f}{\text{arg min}}\text{ }
     (\Delta C_i^{(n)}), \quad \\
     &\Delta C_i^{(n)}=
         C_{{\text{thre}},i}(2^{d}\mathcal{M}_i^{(n)}, \hat{G}_i^{(n)},P_{b}^{\text{target}})-C_{i}^{(n)}
\end{aligned}
\end{equation}
where the subcarrier that requires the smallest incremental power is selected. This greedy process repeats, using the predicted \ac{csi} to recompute required power for each subcarrier, until the power budget is exhausted or no further upgrade satisfies $P_b^{\text{target}}$. If the channel conditions deteriorate, the algorithm downgrades the subcarrier with the largest marginal cost (i.e., set $d=-1$ in \eqref{eq.amc_sub}).

\subsection{Channel Prediction Model}
The multi-step channel prediction problem aims to forecast future \ac{csi} $\{\mathbf{h}^{[t+1]}, \ldots, \mathbf{h}^{[t+T_{\text{pred}}]}\}$ from a historical observation window $\{\mathbf{h}^{[t-T_{\text{win}}+1]}, \ldots, \mathbf{h}^{[t]}\}$, where each $\mathbf{h}^{[t]} \in \mathbb{R}^{N_c}$ is a \ac{csi} vector with $N_c$ features (e.g., multipath taps or subcarrier gains). PatchCSI-T adopts a joint multi-step prediction strategy, generating the entire $T_{\text{pred}}$-step future sequence in a single forward pass to avoid the error accumulation inherent in recursive approaches. 

Let $\mathbf{X} \in \mathbb{R}^{N_c \times T_{\text{win}}}$ denote the input \ac{csi} matrix formed by stacking the historical \ac{csi} vectors over the observation window, where the $k$-th row $\mathbf{x}^{\langle k\rangle} = \mathbf{X}(k,:) \in \mathbb{R}^{T_{\text{win}}}$ represents the temporal sequence of the $k$-th \ac{csi} feature. Instead of processing each time sample independently, PatchCSI-T partitions each feature sequence $\mathbf{x}^{\langle k\rangle}$ into overlapping patches of length $L_{\text{patch}}$ with stride $s_{\text{patch}}$, yielding $N_{\text{patch}} = \lfloor (T_{\text{win}} - L_{\text{patch}})/s_{\text{patch}} \rfloor + 2$ tokens after end-padding. This reduces the effective sequence length from $T_{\text{win}}$ to $N_{\text{patch}}$, lowering the self-attention complexity from $O(T_{\text{win}}^2)$ to $O(N_{\text{patch}}^2)$, while enabling the model to capture temporal dependencies at the segment level rather than between isolated time samples. Each patch is then mapped to a $d_{\text{model}}$-dimensional space via a linear embedding and combined with sinusoidal positional encoding. 

At the core of PatchCSI-T is the Transformer encoder layer which globally models the relationships among all patch tokens. Existing channel prediction methods often adopt feature-mixing modeling to capture inter-feature correlations (e.g., across multipath taps or subcarriers). However, recent studies in time-series forecasting suggest that feature-independent modeling is often more robust \cite{hanCapacityRobustnessTradeOff2024}, especially when inter-feature correlations are weak or non-stationary. Moreover, feature mixing increases model complexity and data requirements, which can aggravate overfitting when training data are limited. PatchCSI-T therefore adopts feature-independent modeling. Specifically, as shown in Fig. \ref{fig:framework}(c), the feature dimension $N_c$ is merged into the batch dimension $B$, yielding $\mathbf{X}\in\mathbb{R}^{(BN_c)\times L_{\text{patch}}\times N_{\text{patch}}}$ after patching. This allows all features to be processed in parallel within a single forward pass. Although features are processed independently, they share the same model parameters and are optimized jointly. Finally, the backbone output is flattened and passed to a linear prediction head that directly generates the full $T_{\text{pred}}$-step future sequence:
\begin{equation}
    \mb{X}_{\text{out}}^{\langle k \rangle}=\mb{W}_P\operatorname{Flatten}(\mb{R}^{\langle k \rangle})+\mb{b}_P
\end{equation}
where $\mb{W}_P\in\mathbb{R}^{T_{\text{pred}}\times (d_{\text{model}}\times N_{\text{patch}})}$, $\mb{b}_P\in\mathbb{R}^{T_{\text{pred}}}$ are learnable parameters.

\section{Numerical Results}
\subsection{Dataset and Experimental Setup}
\begin{figure}[H]
\centering
\subfloat[]{%
    \includegraphics[scale=0.06]{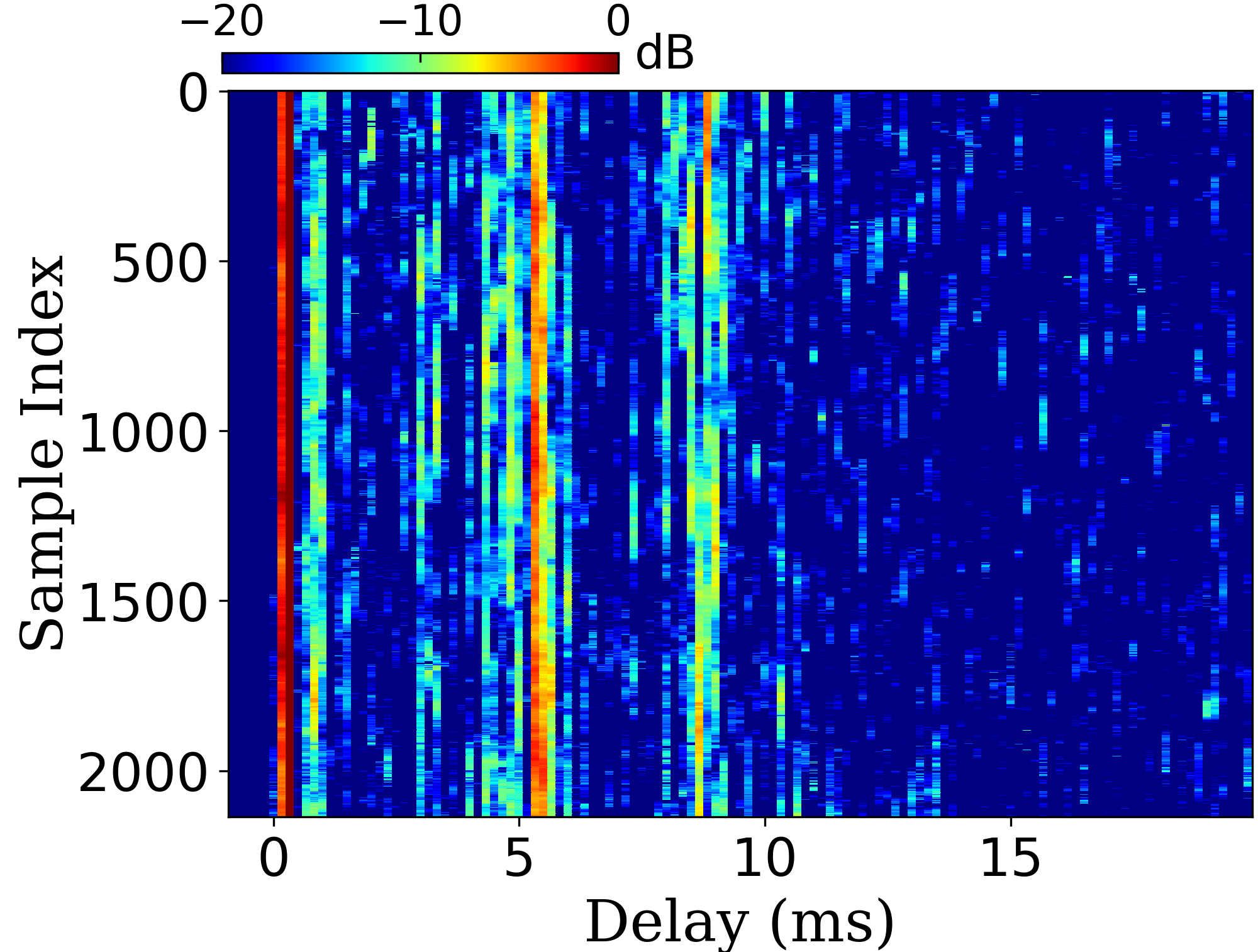}%
    \label{fig.cir_cwr}%
}
\subfloat[]{%
    \includegraphics[scale=0.06]{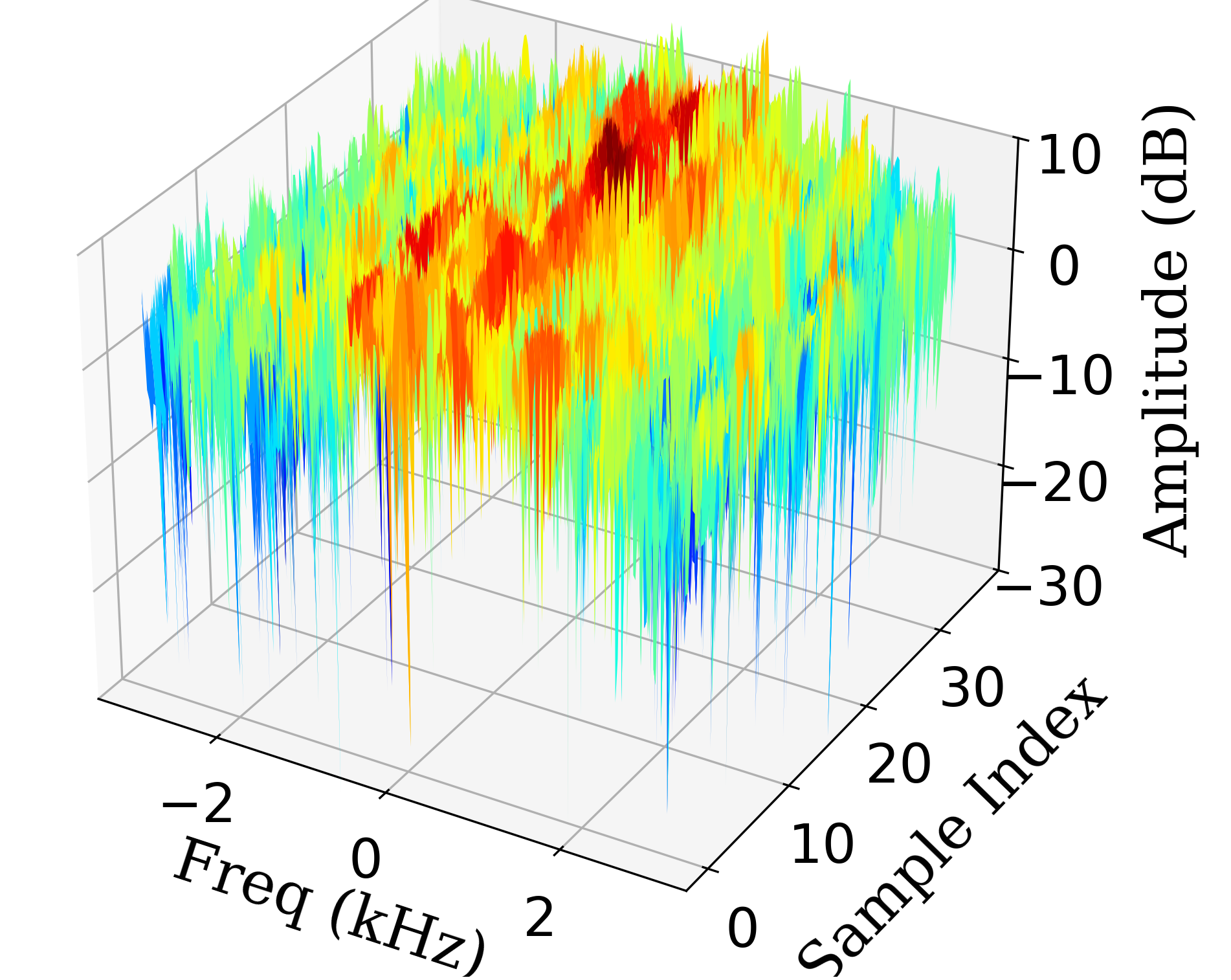}%
    \label{fig.cfr_cwr}%
}
\vspace{-2mm}
\caption{Time-varying channel dataset: (a) \acp{cir}, (b) \acp{cfr}.}
\label{fig.exp_channel}
\end{figure}

We evaluate the performance using the open-access China-Wanlu Reservoir dataset \cite{zhaoFederatedMetaLearningEnhanced2022a}, which includes 2,137 \ac{cir} measurements with maximum delay spread of $\tau_{\max}=$20 ms and $N_{\tau}=120$ multipath taps. To assess frequency-domain prediction performance, we apply Fourier transform to the \ac{cir} data, corresponding to a bandwidth of $BW = N_{\tau}/\tau_{\max} = 6\,\mathrm{kHz}$ and $N_f=256$ subcarrier bins in \ac{cfr}. Fig. \ref{fig.exp_channel} illustrates both the time-varying characteristics of the \ac{cir} and \ac{cfr}. The dataset is partitioned into 70\% training, 10\% validation, and 20\% test sets in chronological order.

For the PatchCSI-T model, the patch length is set to $L_{\text{patch}}=16$, the stride to $s_{\text{patch}}=8$, corresponding to 50\% overlap, and the hidden dimension to $d_{\text{model}}=64$. The model is trained using the \ac{rmse} loss with a historical window of $T_{\text{win}} = 64$. Prediction accuracy is measured by \ac{nmse}:
\begin{equation}
    \text{NMSE} =  \frac{1}{N_s} \sum_{n=1}^{N_s} \frac{\|\hat{\mathbf{y}}^{[n]} - \mathbf{y}^{[n]}\|_2^2}{\|\mathbf{y}^{[n]}\|_2^2} ,
\end{equation}
where $\mathbf{y}^{[n]}$, $\hat{\mathbf{y}}^{[n]}$ $\in \mathbb{R}^{N_f}$ denote the ground-truth and predicted \ac{cfr} vectors corresponding to the $n$-th sample in the test set, and $N_s$ denotes the total number of test samples.

We compare PatchCSI-T against \ac{rls}, BiGRU \cite{huChannelPredictionUsing2023}, CNN1d-LSTM \cite{liuChannelStateInformation2021a}, and MTL-LSTM \cite{tian}. For the \ac{dl} baselines, $d_{\text{model}}$ is selected by hyperparameter search over \{64,128,256\}, resulting in $d_{\text{model}}=64$ for MTL-LSTM and $d_{\text{model}}=256$ for BiGRU and CNN1d-LSTM. All \ac{dl} models are trained under the same protocol using the AdamW optimizer. Experiments are conducted on a workstation equipped with an NVIDIA GeForce RTX 4090 GPU, a 13th Gen Intel Core i9-13900K CPU, and 64 GB RAM.

\subsection{Prediction Performance Analysis}
\begin{figure}[H]
    \centering
    \includegraphics[width=0.52\textwidth]{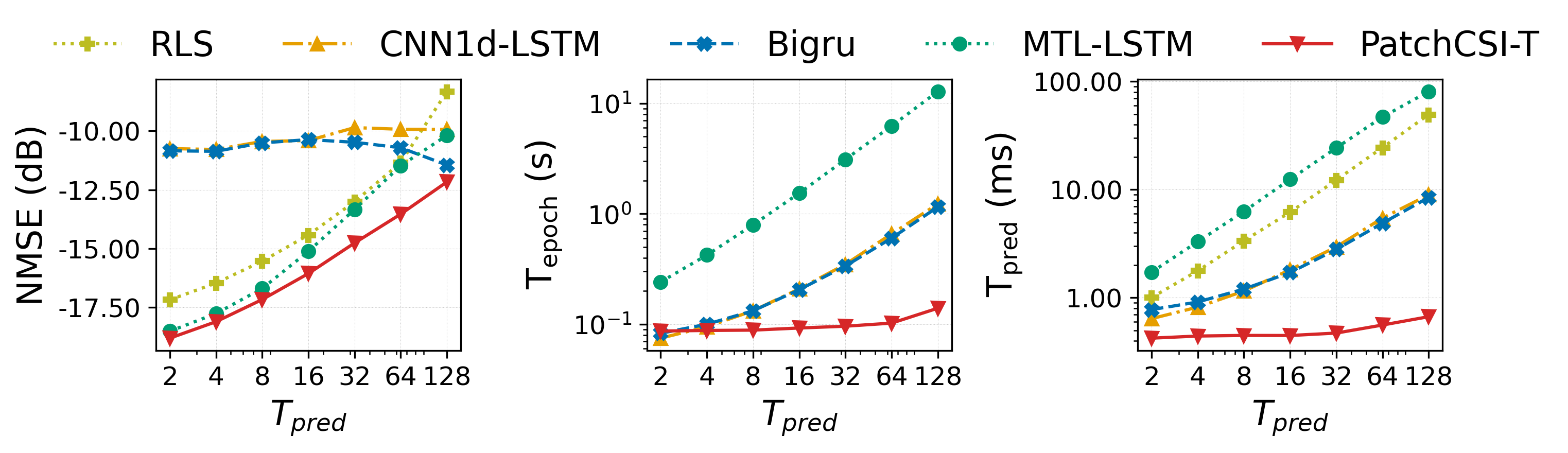}
    \caption{Comparison of prediction performance.}
    \label{fig:pred_curve}
\end{figure}

Fig. \ref{fig:pred_curve} summarizes \ac{cfr} predictions over horizons $T_{\text{pred}}\in\{2,4,8,16,32, 64, 128\}$, including \ac{nmse}, per-epoch training time ($\text{T}_{\text{epoch}}$) and inference time ($\text{T}_{\text{pred}}$). PatchCSI-T consistently achieves higher accuracy and faster inference. At $T_{\text{pred}}=32$, it attains -14.76 dB \ac{nmse}, a 1.4-4.9 dB \ac{nmse} gain over the baselines, with 14.7 ms inference ($\sim$6$\times$ faster than CNN1d-LSTM/BiGRU and $>50\times$ faster than MTL-LSTM). Notably, CNN1d-LSTM and BiGRU exhibit \ac{nmse} saturation near -10 dB for large prediction horizon ${T}_{\text{pred}}$, as these models were originally developed for single-step prediction tasks on larger datasets (e.g., CsiPreNet employs a similar CNN1d–LSTM structure trained on 10,584 samples for 34 subcarrier clusters \cite{liuChannelStateInformation2021a}).

\begin{figure*}
    \centering
    \includegraphics[width=0.99\textwidth]{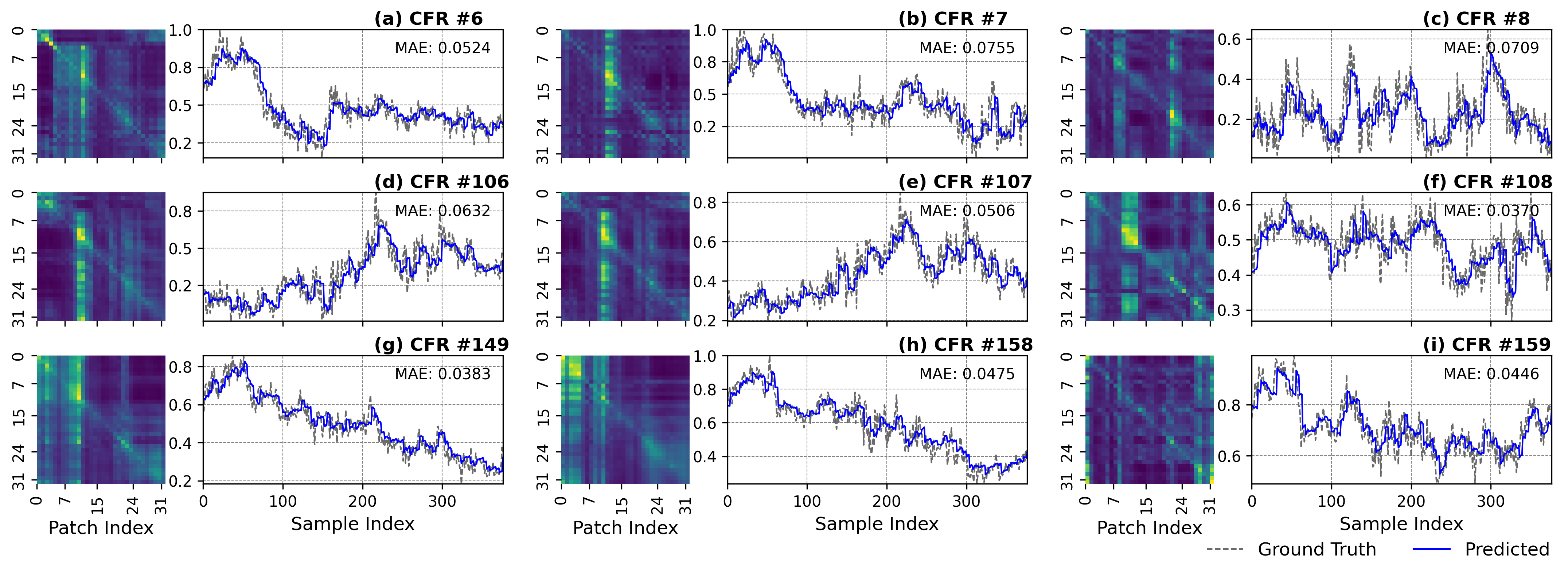}\vspace{-3mm}
    \caption{Multi-head attention weight matrices and prediction curves for selected subcarrier components.}
    \label{fig:attn}
\end{figure*}

To understand how PatchCSI-T captures temporal dependencies, Fig. \ref{fig:attn} shows the multi-head attention weight matrices, averaged over all attention heads, together with the corresponding multi-step prediction curves for selected \ac{cfr} subcarrier features. In this example, the model uses $T_{\text{win}} = 256$ and $T_{\text{pred}} = 4$. The feature-independent strategy enables the model to learn distinct attention patterns for channel components with different temporal characteristics. Features with similar temporal dynamics, such as adjacent subcarriers experiencing correlated fading, exhibit similar attention structures. This indicates that the shared Transformer backbone can capture feature-specific temporal dependencies without explicit cross-feature modeling.
\begin{figure*}
    \centering
    \includegraphics[width=0.99\textwidth]{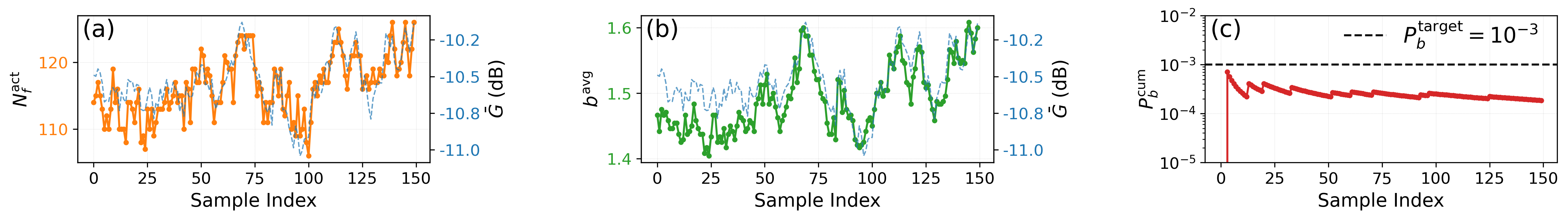}\vspace{-3mm}
    \caption{Performance of the adaptive OFDM scheme under ground-truth \ac{cfr} at SNR=5 dB: (a) number of active subcarriers $N_f^{\mathrm{act}}$ and average channel gain $\bar{G}$; (b) average number of bits per data subcarrier $b^{\mathrm{avg}}$; and (c) cumulative \ac{ber} $P_e^{\mathrm{cum}}$.}
    \label{fig:amc_gt}
\end{figure*}

\subsection{Performance in Adaptive OFDM System}
The OFDM system is configured with $N_f = 256$ subcarriers over a bandwidth of $B_W = 6$ kHz, corresponding to a subcarrier spacing of $\Delta f = 23.4$ Hz. Among them, 240 subcarriers are allocated for data transmission and 16 are null subcarriers reserved for guard bands and noise monitoring. The OFDM symbol duration is $T_{\mathrm{OFDM}} = 42.7$ ms, with a cyclic prefix of $T_{\mathrm{CP}} = 20$ ms. Each data subcarrier is assigned one of five states: deactivated, BPSK, QPSK, 8PSK, or 16QAM. The target bit error rate is set to $P_b^{\mathrm{target}} = 10^{-3}$. Subcarriers that cannot support even BPSK at the target \ac{ber} are deactivated, and their power is reallocated to the remaining active subcarriers.

We first validate the greedy modulation and power allocation algorithm using ground-truth \ac{cfr}. Fig. \ref{fig:amc_gt} shows the temporal evolution of the allocation results together with the average channel gain $\bar{G}$. As shown in Fig. \ref{fig:amc_gt} (a) and (b), both the number of active subcarriers $N_f^{\mathrm{act}}$ and the average number of bits per data subcarrier $b^{\mathrm{avg}}$ closely track the variation of $\bar{G}$: when the average channel gain increases, more subcarriers are activated and higher-order modulation levels are selected. During channel fades, the number of active subcarriers and the modulation order are reduced to maintain the target \ac{ber}. Fig. \ref{fig:amc_gt} (c) shows that the cumulative \ac{ber} remains well below $10^{-3}$, confirming that the adaptive algorithm can maintain reliable transmission under time-varying channel conditions.

\begin{figure}[H]
    \centering
    \includegraphics[width=0.48\textwidth]{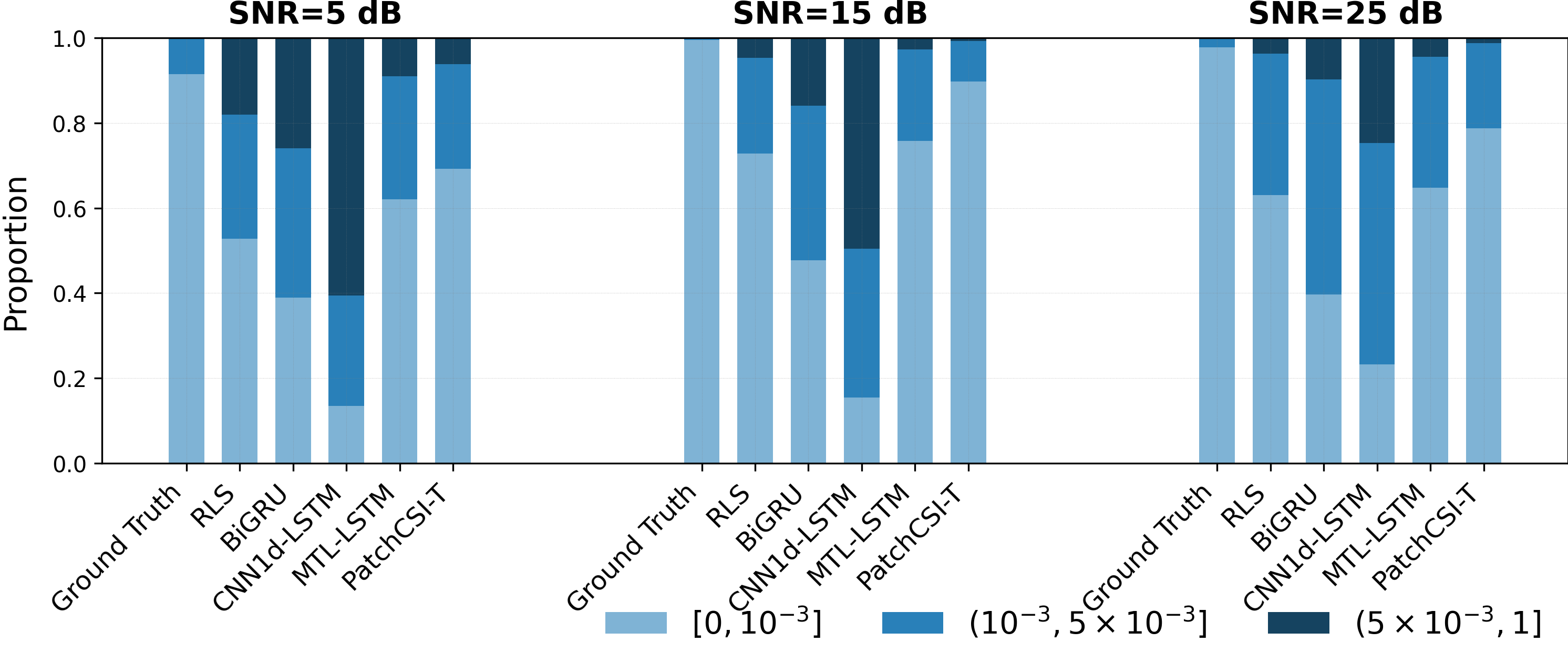}
    \caption{BER distribution under real and predicted \acp{cfr}.}
    \label{fig:amc_ber_distribution}
\end{figure}

\begin{figure}[H]
    \centering
    \includegraphics[width=0.48\textwidth]{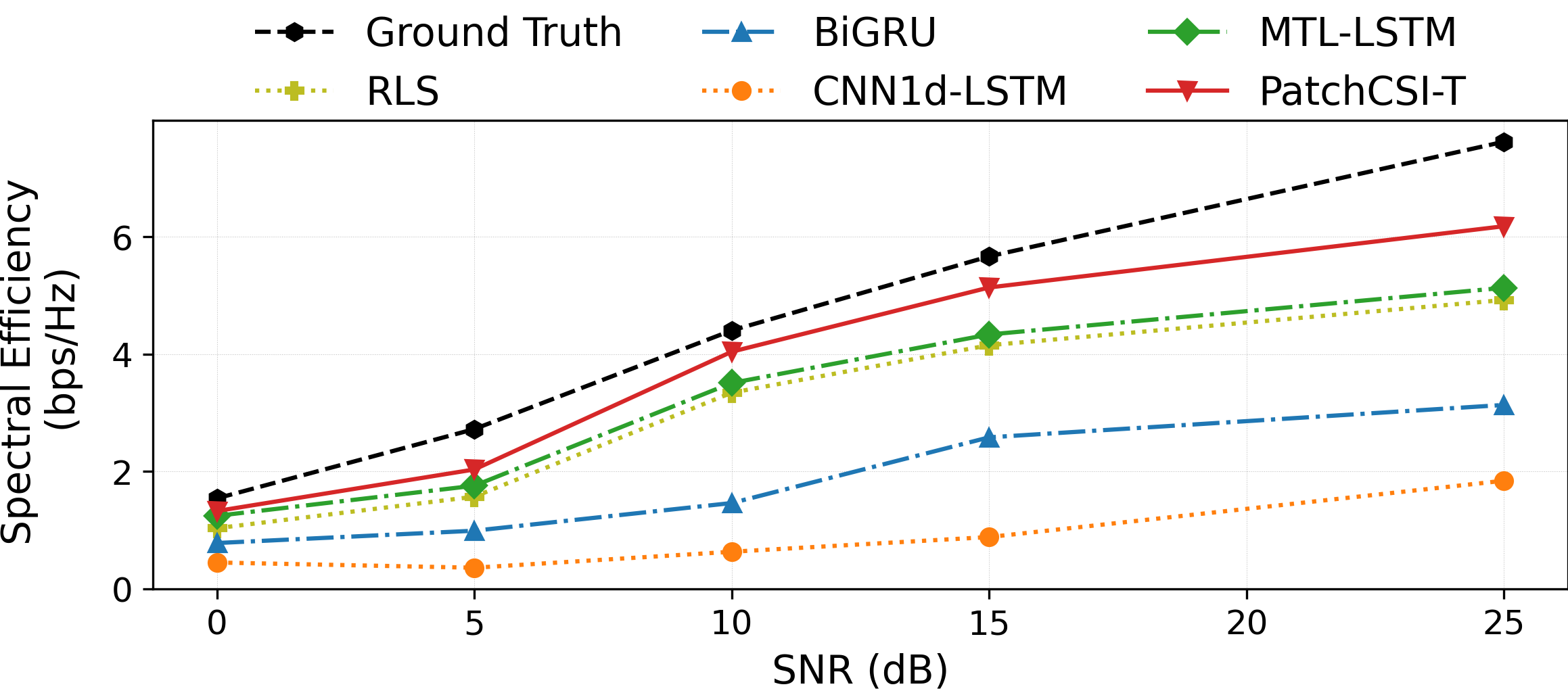}
    \caption{Spectral efficiency for the adaptive OFDM system.}
    \label{fig:amc_throughput}
\end{figure}

We next evaluate the adaptive OFDM system under predicted \ac{cfr}. The channel prediction horizon is $T_{\mathrm{pred}} = 32$. Fig. \ref{fig:amc_ber_distribution} shows the \ac{ber} distribution across three \ac{snr} levels using both ground-truth and predicted \ac{cfr} from each model. Fig. \ref{fig:amc_throughput} plots the effective spectral efficiency (SE) defined as
{\small{
\begin{equation}
    SE=\frac{\sum_{n=1}^{N_{s}}\sum_{i=1}^{N_f}I(P_b^{(n)}<P_b^{(\text{target})})\log_2\mathcal{M}_i^{(n)}}{N_s\times BW\times (T_{\text{OFDM}}+T_{\text{CP}})}.
\end{equation}}}
Across all \ac{snr} levels, PatchCSI-T achieves performance closest to ground truth \ac{csi}. For instance, at SNR=15dB, PatchCSI-T maintains 89.7\% of transmitted symbols with $P_b^{\text{target}}\leq10^{-3}$ and 5.14 bps/Hz spectral efficiency, outperforming RLS (72.9\%, 4.15 bps/Hz) and MTL-LSTM (75.8\%, 4.34 bps/Hz). 

\section{Conclusions}
This paper presents PatchCSI-T, a joint multi-step \ac{csi} prediction model for adaptive OFDM in time-varying \ac{uwa} channels. Results on the open-access China-Wanlu Reservoir dataset show that PatchCSI-T improves prediction accuracy by 1.4–4.9 dB in \ac{nmse} at horizon $T_{\mathrm{pred}}=32$ over the baselines, while reducing inference time by more than $50\times$ compared with MTL-LSTM. Integrated with a greedy modulation and power allocation scheme, the proposed framework achieves 5.14 bps/Hz spectral efficiency at SNR = 15 dB, with 89.7\% of transmitted symbols satisfying the target \ac{ber}. Future work will extend validation to at-sea data with stronger Doppler and longer delays, and to real-time closed-loop experiments.

\section{Acknowledgment}
This work was supported in part by the National Natural Science Foundation of China under Grants 42306207 and 62171369, in part by the National Key Research and Development Program of China under Grant 2024YFF0510000, in part by the Natural Science Foundation of Fujian Province under Grant 2025J01860, and in part by the Research Ireland Sea-Scan project under grant 24/FIP/DO/13340P and 13/RC/2106\_P2 (ADAPT centre) and 18/RI/5721 (OpenIreland Research Infrastructure).
\begin{acronym}
  \acro{ser}[SER]{Symbol Error Rate}
  \acro{uacs}[UACs]{Underwater Acoustic Channels}
  \acro{uac}[UAC]{Underwater Acoustic Channel}
  \acro{uwa}[UWA]{Underwater Acoustic}
  \acro{ti}[TI]{time-invariant}
  \acro{dd}[DD]{Delay-Doppler}
  \acro{ici}[ICI]{Intercarrier Interference}
  \acro{cir}[CIR]{Channel Impulse Response}
  \acro{cp}[CP]{Cyclic Prefix}
  \acro{dft}[DFT]{Discrete Fourier Transform}
  \acro{ofdm}[OFDM]{Orthogonal Frequency Division Multiplexing }
  \acro{ipm}[IPM]{Interior-Point Method}
  \acro{admm}[ADMM]{Alternating Direction Method of Multipliers}
  \acro{pgm}[PGM]{Proximal Gradient Method}
  \acro{apg}[APG]{Accelerated Proximal Gradient}
  \acro{alf}[ALF]{Augmented Lagrangian Function}
  \acro{alm}[ALM]{Augmented Lagrangian Method}
  \acro{omp}[OMP]{Orthogonal Matching Pursuit}
  \acro{fista}[FISTA]{Fast Iterative Shrinkage-Thresholding Algorithm}
  \acro{ampa}[AMPA]{Adaptive Modulation and Power Allocation}
  
  \acro{snr}[SNR]{Signal-to-Noise Ratio}
  \acro{nmsd}[NMSD]{Normalized Mean-Square Deviation}
  \acro{wgn}[WGN]{White Gaussian Noise}
  \acro{gmn}[GMN]{Gaussian Mixture Noise}
  \acro{cs}[CS]{Compressed Sensing}
  \acro{rip}[RIP]{Restricted Isometry Property}
  \acro{caf}[CAF]{Cross-Ambiguity Function}
  \acro{mmse}[MMSE]{Minimum Mean Squared Error }
  \acro{ber}[BER]{Bit Error Rate}
  \acro{ser}[SER]{Symbol Error Rate}
  \acro{prbs}[PRBS]{Pseudo-Random Binary Sequence}
  \acro{bpsk}[BPSK]{Binary Phase-Shift Keying}
  \acro{qpsk}[QPSK]{Quadrature Phase Shift Keying}
  \acro{srrc}[SRRC]{Square Root Raised Cosine}
  \acro{caf}[CAF]{Cross-Ambiguity Function}
  \acro{inr}[INR]{Interference-to-Noise Ratio}
  \acro{awgn}[AWGN]{Additive White Gaussian Noise}
  \acro{sinr}[SINR]{Signal-to-Interference-plus-Noise Ratio}

  \acro{ls}[LS]{Least Squares}
  \acro{pn}[PN]{Pseudorandom Noise}
  \acro{mimo}[MIMO]{Multiple-Input and Multiple-Output}
  \acro{uwa}[UWA]{Underwater Acoustic}
  \acro{rt}[RT]{Real Time}
  \acro{ric}[RIC]{RAN Intelligent Controller}
  \acro{mac}[MAC]{Medium Access Control}
  \acro{bs}[BS]{Base Station}
  \acro{ue}[UE]{User Equipment}
  \acro{cqi}[CQI]{Channel Quality Indicator}
  \acro{mcs}[MCS]{Modulation and Coding Scheme}
  \acro{pusch}[PUSCH]{Physical Uplink Shared Channel}
  \acro{pucch}[PUCCH]{Physical Uplink Control Channel}
  \acro{usrp}[USRP]{Universal Software Radio Peripheral}  \acro{vm}[VM]{Virtual Machine}
  \acro{voip}[VoIP]{Voice over IP}
  \acro{dos}[DoS]{Denial-of-Service}
  \acro{tcp}[TCP]{Transmission Control Protocol}
  \acro{qos}[QoS]{Quality of Service}
  \acro{udp}[UDP]{User Datagram Protocol}
  \acro{svm}[SVM]{Support Vector Machine}
  \acro{knn}[k-NN]{k-Nearest Neighbors}
  \acro{adaboost}[AdaBoost]{Adaptive Boosting}
  \acro{mlp}[MLP]{Multilayer Perceptron}
  \acro{ids}[IDS]{Intrusion Detection System}
  \acro{du}[DU]{Distributed Unit}
  \acro{cu}[CU]{Centralized Unit}
  \acro{rrc}[RRC]{Radio Resource Control}
  \acro{dpi}[DPI]{Deep Packet Inspection}
  \acro{csi}[CSI]{Channel State Information}
  \acro{cfr}[CFR]{Channel Frequency Response}
  \acro{le}[LE]{Linear Embedding}
  \acro{mhsa}[MHSA]{Multi-Head Self-Attention}
  \acro{ff}[FF]{Feedforward}
  \acro{rmse}[RMSE]{Root Mean-Square Error}
  \acro{dl}[DL]{Deep Learning}
  \acro{gru}[GRU]{Gated Recurrent Unit}
  \acro{lstm}[LSTM]{Long Short-Term Memory}
  \acro{nmse}[NMSE]{Normalized Mean-Square Error}
  \acro{nn}[NN]{Neural Network}
  \acro{rnn}[NN]{Recurrent Neural Network}
  \acro{rls}[RLS]{Recursive Least Squares}
  \acrodef{mtl}[MTL]{Multi-Task Learning}
\end{acronym}

\bibliographystyle{IEEEtran}
\bibliography{IEEEabrv, Oceans2025}
\end{document}
